\documentclass[conference]{IEEEtran}
\IEEEoverridecommandlockouts
\usepackage[T1]{fontenc}
\usepackage[utf8]{inputenc}
\usepackage{cite}
\usepackage{amsmath,amssymb,amsfonts}
\usepackage{algorithmic}
\usepackage{graphicx}
\usepackage{textcomp}
\usepackage{xcolor}
\usepackage{url}
\usepackage[english]{babel}
\usepackage{microtype}
\usepackage[free-standing-units,per-mode=repeated-symbol,mode=text,binary-units=true,detect-weight=true,detect-family=true]{siunitx}
\usepackage{pgfplots}
\usepackage{acronym}
\usepackage{tikz}
\usepackage{booktabs}
\usepackage{glossaries}
\usepackage{cleveref}
\usepackage{pifont}
\usepackage{microtype}
\usepackage{breakurl}

\newcommand{\cmark}{\ding{51}}%
\newcommand{\xmark}{\ding{55}}%
\usepackage[para,online,flushleft]{threeparttable}

\renewcommand{\baselinestretch}{1}

\usetikzlibrary{scopes, calc, shapes, arrows, positioning}
\tikzset{>=latex}
\glsdisablehyper

\pgfplotsset{compat=1.13}
\pgfplotsset{width=\linewidth, height=7cm}
\pgfplotsset{every x tick label/.append style={font=\small}}
\pgfplotsset{every y tick label/.append style={font=\small}}

\pgfplotsset{
  /pgf/declare function={
    roof(\x,\b,\p) = (\b * \x < \p) * \b * \x + (\b * \x >= \p) * \p;}}

\usepgfplotslibrary{external}
\usepgfplotslibrary{groupplots}
\usepgfplotslibrary{fillbetween}

\pgfplotsset{compat=1.13}
\pgfplotsset{width=\linewidth, height=7cm}
\pgfplotsset{every x tick label/.append style={font=\small}}
\pgfplotsset{every y tick label/.append style={font=\small}}

\DeclareSIUnit\core{core}
\DeclareSIUnit\cycle{cycle}
\DeclareSIUnit\flop{FLOP}
\DeclareSIUnit\flops{FLOPS}
\DeclareSIUnit\gate{GE}
\DeclareSIUnit\op{OP}
\DeclareSIUnit\ops{OPS}
\DeclareSIPrefix\double{DP-}{0}
\DeclareSIPrefix\dpmega{DP-M}{6}
\DeclareSIPrefix\dpgiga{DP-G}{9}
\DeclareSIPrefix\dptera{DP-T}{12}
\definecolor{color1}{HTML}{256DFF}
\definecolor{color2}{HTML}{45CCB0}
\definecolor{color3}{HTML}{9775CA}
\definecolor{color4}{HTML}{C83737}
\definecolor{color5}{HTML}{000000}
\definecolor{colorAlert}{HTML}{FF0000}

\newcommand\new[1]{\textcolor{black}{#1}}
\newcommand\neww[1]{\textcolor{black}{#1}}
\newcommand\camrdy[1]{\textcolor{black}{#1}}

\newcommand\eg{e.g.,\ }
\newcommand\ie{i.e.,\ }
\newcommand\etal{et\penalty50\ al.\ }

\newcommand\eew[2]{$\text{EEW}_{#1}^{\text{#2}}$}
\newcommand{\tabref}[1]{Table~\ref{#1}}

\newacronym[longplural={Scratchpad Memories}]{SPM}{SPM}{Scratchpad Memory}
\newacronym{ACE}{ACE}{AXI Coherent Extensions}
\newacronym{AMBA}{AMBA}{Advanced Microcontroller Bus Architecture}
\newacronym{APB}{APB}{Advanced Peripheral Bus}
\newacronym{API}{API}{Application Programming Interface}
\newacronym{ASIC}{ASIC}{Application-Specific Integrated Circuit}
\newacronym{AVX}{AVX}{Advanced Vector Extension}
\newacronym{AXI}{AXI}{Advanced eXtensible Interface}
\newacronym{BLAS}{BLAS}{Basic Linear Algebra Subprograms}
\newacronym{CHI}{CHI}{Coherent Hub Interface}
\newacronym{CMOS}{CMOS}{Complementary Metal-Oxide-Semiconductor}
\newacronym{CNN}{CNN}{Convolutional Neural Network}
\newacronym{CPU}{CPU}{Central Processing Unit}
\newacronym{CSR}{CSR}{Control and State Register}
\newacronym{CTS}{CTS}{Clock Tree Synthesis}
\newacronym{DLP}{DLP}{Data Level Parallelism}
\newacronym{DMA}{DMA}{Direct Memory Access}
\newacronym{DRAM}{DRAM}{Dynamic Random-Access Memory}
\newacronym{DSP}{DSP}{Digital Signal Processing}
\newacronym{DUT}{DUT}{Device Under Test}
\newacronym{ECL}{ECL}{Emitter-Coupled Logic}
\newacronym{FBB}{FBB}{Forward Body-Biasing}
\newacronym{FDSOI}{FD-SOI}{Fully Depleted Silicon-on-Insulator}
\newacronym{FMA}{FMA}{Fused Multiply-Add}
\newacronym{FPGA}{FPGA}{Field-Programmable Gate Array}
\newacronym{FPU}{FPU}{Floating Point Unit}
\newacronym{GPGPU}{GPGPU}{General-Purpose \acrlong{GPU}}
\newacronym{GPU}{GPU}{Graphics Processing Unit}
\newacronym{HDL}{HDL}{Hardware Description Language}
\newacronym{HERO}{HERO}{Heterogeneous Embedded Research Platform}
\newacronym{HPC}{HPC}{High-Performance Computing}
\newacronym{ILP}{ILP}{Instruction Level Parallelism}
\newacronym{IOT}{IoT}{Internet-of-Things}
\newacronym{IPC}{IPC}{Instructions Per Cycle}
\newacronym{IPU}{IPU}{Image Processing Unit}
\newacronym{ISA}{ISA}{Instruction Set Architecture}
\newacronym{LSB}{LSB}{Least Significant Bit}
\newacronym{LSU}{LSU}{Load/Store Unit}
\newacronym{LVT}{LVT}{low voltage threshold}
\newacronym{MIMD}{MIMD}{multiple instruction, multiple data}
\newacronym{MMU}{MMU}{Memory Management Unit}
\newacronym{MUL}{MUL}{multiplier}
\newacronym{ML}{ML}{Machine Learning}
\newacronym{MVL}{MVL}{maximum vector length}
\newacronym{NUMA}{NUMA}{non-uniform memory access}
\newacronym{NOC}{NoC}{Network-on-Chip}
\newacronym{PCIe}{PCIe}{Peripheral Component Interconnect Express}
\newacronym{PC}{PC}{Program Counter}
\newacronym{PE}{PE}{processing element}
\newacronym{PL}{PL}{Programmable Logic}
\newacronym{PMCA}{PMCA}{Programmable Manycore Accelerator}
\newacronym{PSL}{PSL}{Power Service Layer}
\newacronym{PTE}{PTE}{page-table entry}
\newacronym{PTW}{PTW}{page-table walker}
\newacronym{PULP}{PULP}{Parallel Ultra Low Power}
\newacronym{RAW}{RAW}{read-after-write}
\newacronym{RBB}{RBB}{Reverse Body-Biasing}
\newacronym{ROB}{ROB}{Reorder Buffer}
\newacronym{RTL}{RTL}{Register Transfer Level}
\newacronym{RVT}{RVT}{Regular Voltage Threshold}
\newacronym{RoCC}{RoCC}{Rocket Custom Coprocessor Interface}
\newacronym{SCM}{SCM}{Storage Class Memory}
\newacronym{SIMD}{SIMD}{single instruction, multiple data}
\newacronym{SIMT}{SIMT}{single instruction, multiple thread}
\newacronym{SLDU}{SLDU}{Slide Unit}
\newacronym{SLVT}{SLVT}{super-low voltage threshold}
\newacronym{SM}{SM}{Streaming Multiprocessor}
\newacronym[longplural={Static Random-Access Memories}]{SRAM}{SRAM}{Static Random-Access Memory}
\newacronym{SSE}{SSE}{Streaming SIMD Extension}
\newacronym{SVE}{SVE}{Scalable Vector Extension}
\newacronym{TLP}{TLP}{Thread Level Parallelism}
\newacronym{TxnID}{TxnID}{Transaction ID}
\newacronym{VAC}{VAC}{Vector Access}
\newacronym{VC}{VC}{virtual channel}
\newacronym{VCONV}{VCONV}{Vector Conversion}
\newacronym{VEX}{VEX}{Vector Execute}
\newacronym{VFU}{VFU}{vector functional unit}
\newacronym{VID}{VID}{Vector Instruction Decode}
\newacronym{VIS}{VISSUE}{Vector Instruction Issue}
\newacronym{VLIW}{VLIW}{Very Long Instruction Word}
\newacronym{VLOOP}{VLOOP}{Vector Loop}
\newacronym{VLR}{VLR}{vector length register}
\newacronym{VLSU}{VLSU}{Vector Load/Store Unit}
\newacronym{VNB}{VNB}{Von Neumann Bottleneck}
\newacronym{VRF}{VRF}{Vector Register File}
\newacronym{VT}{VT}{vector thread}
\newacronym{MASKU}{MASKU}{Mask Unit}
\newacronym{VU0.5}{VU0.5}{Vector Unit 0.5}
\newacronym{VU1.0}{VU1.0}{Vector Unit 1.0}
\newacronym{VMFPU}{VMFPU}{Vector Multiplier/Floating Point Unit}
\newacronym{VFPU}{VFPU}{Vector Floating Point Unit}
\newacronym{VDIV}{VDIV}{Vector Divider}
\newacronym{VMUL}{VMUL}{Vector Multiplier}
\newacronym{WAR}{WAR}{write-after-read}
\newacronym{WAW}{WAW}{write-after-write}
\newacronym{DCT}{DCT}{discrete cosine transform}
\newacronym{TSV}{TSV}{through-silicon via}
\newacronym{3DIC}{3D-IC}{three-dimensional integrated circuit}
\newacronym{PPA}{PPA}{power, performance, and area}
\newacronym{F2F}{F2F}{face-to-face}
\newacronym{IC}{IC}{integrated circuit}
\newacronym{C4}{C4}{controlled collapse chip connection}
\newacronym{FEOL}{FEOL}{front end of the line}
\newacronym{BEOL}{BEOL}{back end of the line}
\newacronym{SLEN}{SLEN}{striping distance}
\newacronym{RVV}{RVV}{RISC-V Vector Extension}

\begin{document}

\title{A ``New Ara'' for Vector Computing: An Open Source Highly Efficient RISC-V V 1.0 Vector Processor Design
\thanks{\textcopyright 2022 IEEE.  Personal use of this material is permitted.  Permission from IEEE must be obtained for all other uses, in any current or future media, including reprinting/republishing this material for advertising or promotional purposes, creating new collective works, for resale or redistribution to servers or lists, or reuse of any copyrighted component of this work in other works.}
\thanks{}
\thanks{DOI: 10.1109/ASAP54787.2022.00017}
}

\author{\IEEEauthorblockN{Matteo Perotti}
\IEEEauthorblockA{\textit{Integrated Systems Laboratory} \\
\textit{ETH Zurich}\\
Zurich, Switzerland \\
mperotti@iis.ee.ethz.ch}
\and
\IEEEauthorblockN{Matheus Cavalcante}
\IEEEauthorblockA{\textit{Integrated Systems Laboratory} \\
\textit{ETH Zurich}\\
Zurich, Switzerland \\
matheusd@iis.ee.ethz.ch}
\and
\IEEEauthorblockN{Nils Wistoff}
\IEEEauthorblockA{\textit{Integrated Systems Laboratory} \\
\textit{ETH Zurich}\\
Zurich, Switzerland \\
nwistoff@iis.ee.ethz.ch}
\and
\IEEEauthorblockN{Renzo Andri}
\IEEEauthorblockA{\textit{Computing Systems Laboratory} \\
\textit{Huawei Zurich Research Center}\\
Zurich, Switzerland \\
renzo.andri@huawei.com}
\and
\IEEEauthorblockN{Lukas Cavigelli}
\IEEEauthorblockA{\textit{Computing Systems Laboratory} \\
\textit{Huawei Zurich Research Center}\\
Zurich, Switzerland \\
lukas.cavigelli@huawei.com}
\and
\IEEEauthorblockN{Luca Benini}
\IEEEauthorblockA{\textit{Integrated Systems Laboratory} \\
\textit{ETH Zurich and University of Bologna}\\
Zurich, Switzerland \\
lbenini@iis.ee.ethz.ch}
}

\maketitle
\begin{abstract}
Vector architectures are gaining traction for highly efficient processing of data-parallel workloads, driven by all major ISAs (RISC-V, Arm, Intel), and boosted by landmark chips, like the Arm SVE-based Fujitsu A64FX, powering the TOP500 leader Fugaku. The RISC-V V extension has recently reached 1.0-Frozen status. Here, we present its first open-source implementation, discuss the new specification's impact on the micro-architecture of a lane-based design, and provide insights on performance-oriented design of coupled scalar-vector processors. Our system achieves comparable/better PPA than state-of-the-art vector engines that implement older RVV versions: 15\% better area, 6\% improved throughput, and FPU utilization >98.5\% on crucial kernels.
\end{abstract}


\section{Introduction}\label{sec:introduction}

The top places of the current ``Top500'' list of the world's most powerful supercomputers can be divided into two groups: systems drawing their performance from accelerators and systems based on many-core processors with powerful vector units. The current leader is Fujitsu's Fugaku~\cite{sato2020co, dongarra2020report} at the RIKEN research lab in Japan with 159 thousand nodes, each with a 48-core Fujitsu A64FX running at \SI{2.2}{\giga\hertz}, supporting the Armv8.2-A \gls{ISA} with \gls{SVE} using \SI{512}{\bit} vectors. The Sunway TaihuLight~\cite{gao2021sunway} similarly builds on processors with vector extension---and while it is now ranked as number 4 (as of September 2021), it has been on the list since 2016. It has 26 thousand Sunway SW26010 CPUs, each chip running at \SI{1.45}{\giga\hertz} with 4 clusters of 64 compute cores with a 256-bit vector unit each. 

Vector processors find use not only in supercomputers. Arm's latest v9 \gls{ISA} also includes the revised \gls{SVE}2 vector extension \cite{arm-v9} and is expected to become widely used in everyday devices such as smartphones and later on also make it into microcontrollers, real-time, and application processors. 

The aforementioned \glspl{ISA} are all proprietary, leaving their micro-architectural implications completely opaque and preventing open-source implementations. This is a significant obstacle to open innovation. Over the last few years, RISC-V has become a well-established, modern, and openly available alternative to proprietary \glspl{ISA}. This has spurred a wave of publicly available implementations and enabled a public discourse on novel, custom \gls{ISA} extensions as well as their micro-architectural implications and their potential standardization. 

In this work, we focus on the \gls{RVV}, initially proposed in 2015~\cite{rvv-0.1}. Over the past six years, it has been intensively discussed, refined, and is now frozen and open for public review, version v1.0~\cite{rvv-1.0}.
The ratification of an extension is a key process for the RISC-V community, as both hardware and software will rely on a stable and standardized, but still open, \gls{ISA}.
 
Vector instructions operate on variable-sized vectors, whose length and element size can be set at runtime. The architecture exploits the application parallelism through long, deeply pipelined datapaths and \gls{SIMD} computation in each functional unit. Moreover, a single vector instruction triggers the calculation of all the elements of a vector.
This was initially introduced in Cray supercomputers and is sometimes called Cray-style vector processing. The benefits of this approach are a typically smaller code size and avoiding fetching and decoding the same instructions repeatedly, with a reduction in I-cache transfers and an increased system efficiency. Furthermore, it improves code portability.
Today, Cray-like vector processors have gained new traction thanks to the race towards energy-efficient designs and the need for computing highly-parallel workloads. Moreover, thanks to their flexibility and straightforward programming model, they became a concrete alternative to \glspl{GPU} when operating on long vectors.
Note that both the Cray-like and SIMD-style extensions are often referred to as vector extensions/instructions, \eg Intel's \gls{AVX} (introduced in 2011 and further developed into AVX2 in 2013 and into AVX-512 in 2016) and Arm's NEON extensions. However, these instructions with fixed-size operands are not vector extensions according to the definition above. 

This work presents the first open-source implementation, including hardware and software, of the \gls{RVV} 1.0 \gls{ISA}, and makes the following contributions:
\begin{enumerate}
\item It describes a design of the \gls{RVV} 1.0 vector unit as a tightly coupled extension unit for the open-source CVA6 RV64GC processor. It introduces a new interface and hardware memory coherency/consistency features.
\item It presents a comparison with a \gls{RVV} 0.5 vector unit \cite{Ara2020} and an analysis of the impact that the new \gls{RVV} \gls{ISA} has on microarchitectures that use lanes with a split vector register file.
\item It provides a \gls{PPA} evaluation of the unit implemented with \textsc{GlobalFoundries} 22FDX \gls{FDSOI} technology, proving that the implementation achieves competitive \gls{PPA} compared to the \gls{RVV} 0.5 vector unit.
\item It analyzes the impact that the scalar processor has on the final achieved vector throughput, especially in the case of medium/short vectors. 
\end{enumerate}


\section{Related Work}
\label{sec:relWork}

\begin{table}
\caption{Overview of RISC-V Vector Processors}
\label{tab:relwork}
\scriptsize
\begin{threeparttable}
\begin{tabular}{@{}r@{\hspace{1mm}}l@{\hspace{1mm}}c@{\hspace{1mm}}c@{\hspace{1mm}}c@{\hspace{1mm}}c@{\hspace{1mm}}c@{\hspace{1mm}}c@{\hspace{1mm}}c@{\hspace{1mm}}c@{\hspace{1mm}}c@{}}
\toprule
 &\textbf{Core Name} & \textbf{RVV} & \textbf{Target} & \textbf{XLEN}  & \textbf{float} & \textbf{VLEN} & \textbf{Split VRF}  & \textbf{Open-}              \\
 & & \textbf{version} &  & \textbf{(bit)}    & \textbf{supp.} &  \textbf{(bit)} & \textbf{(lanes)}  & \textbf{Source}              \\ \midrule
 & This work & 1.0 & ASIC & 64 & \cmark & 4096\tnote{a} & \cmark & \cmark \\
\cite{sifivex280} & \textbf{SiFive X280} & 1.0 & ASIC & 64 & \cmark & 512 & \textbf{?} & \xmark \\ 
\cite{sifivep270} & \textbf{SiFive P270} & 1.0rc & ASIC & 64 & \cmark & 256 & \textbf{?} & \xmark \\ 
\cite{andesnx27v} & \textbf{Andes NX27V}  & 1.0  & ASIC & 64 & \cmark & 512 & \textbf{?} & \xmark \\ 
\cite{atrevido} & \textbf{Atrevido 220} & 1.0 & ASIC & 64 & \cmark & 128-4096 & \cmark & \xmark \\ 
\cite{platzer2021vicuna} & \textbf{Vicuna}       & 0.10 & FPGA & 32 & \xmark & 128-2048 & \xmark & \cmark \\ 

\cite{assir2021arrow} & \textbf{Arrow}        & 0.9  & FPGA & 32 & \xmark  & \textbf{?} & \cmark \tnote{c} & \xmark \\ 

\cite{johns2020minimal} & \textbf{Johns \etal} & 0.8        & FPGA  & 32  & \xmark  & 32 & \xmark & \xmark \\ 

\cite{vitruvius} & \textbf{Vitruvius} & 0.7.1  & ASIC    & 64  & \cmark & 16384 &  \cmark & \xmark  \\
\cite{9138983} & \textbf{XuanTie 910} & 0.7.1  & ASIC    & 64  & \cmark & 128\tnote{b} &  \cmark & \xmark\tnote{d}  \\ 

\cite{risc-v-squared} & \textbf{RISC-}$\mathbf{V^2}$ & \textbf{?} & ASIC & \textbf{?} &  \xmark & 256 & \xmark & \cmark \tnote{e} \\ 

\cite{Ara2020} & \textbf{Ara} & 0.5       & ASIC    & 64  & \cmark & 4096 \tnote{b} &  \cmark & \cmark \\ 

\cite{hwachav5} & \textbf{Hwacha} & Non-Std. & ASIC       & 64  & \cmark & 512\tnote{b} &  \cmark & \cmark \\ 
\bottomrule
\end{tabular}
\begin{tablenotes}
\item[a] Parametric VLEN. In this work, we selected 4096. \item[b] VLEN per lane. \item[c] VRF is split horizontally. \item[d] The vector unit is not open-source. \item[e] The scalar core is not open-source. 
\end{tablenotes}
\end{threeparttable}
\end{table}

The new \gls{RVV} 1.0 vector unit is inspired by Ara~\cite{Ara2020}, a vector unit implementing \gls{RVV} v0.5. Ara is a highly-efficient vector coprocessor developed in 2019 that works in tandem with the application-class processor CVA6 (formerly Ariane)~\cite{8777130} and is compliant with one of the first and loosely defined \gls{RVV} proposals~\cite{rvv-0.6}. Ara achieved \SI{97}{\percent} peak utilization with $256 \times 256$ matrix multiplication with 16 lanes, 33 DP-GFLOPS of throughput, and 41 DP-GFLOPS/W of energy efficiency at more than \SI{1}{\giga\hertz} in typical conditions in a \SI{22}{\nano\meter} node.

In addition to Ara, we provide an overview of RISC-V processors that currently implement a vector extension in \tabref{tab:relwork}. Notably, most of them are limited to a very short VLEN, which avoids many of the challenges, particularly those introduced in the RVV v1.0 specifications and addressed in this work. In terms of performance, the SiFive P270 is claimed to achieve 5.75 CoreMark/MHz, 3.25 DMIPS/MHz, and a 4.6 result on SPECint 2006 \cite{sifivep270}. The XuanTie 910 obtains 7.1 CoreMark/MHz and 6.11/GHz on SPECint 2006. Even though detailed comparisons with other cores than Ara is not possible because they either do not comply with \gls{RVV} (Hwacha) or they are not (fully) released, we can compare the available achieved FPU utilization during computational kernels: SiFive's X280 and Vicuna claim >90\%, and Hwacha obtained >95\%. Our new architecture reaches >98\% of utilization and >35 DP-GFLOPS/W.


\section{The Evolution of RISC-V V}

The RISC-V V extension enables the processing of multiple data using a single instruction, following the computational paradigm of the original Cray vector processor.
This extension introduces 32 registers organized in a \gls{VRF}, with each register storing a set of data elements of the same type (\eg FP32). Typical vector operations work element-wise on two vectors, \ie on elements with the same index. In addition, predicated computation is supported, preventing the processing of some of the elements based on a Boolean mask vector.

The initial RVV extension was proposed in 2015 \cite{rvv-0.1}, with several updates presented by Krste Asanovi\'c and Roger Espasa~\cite{rvv-0.2, rvv-0.5, rvv-0.6} until 2018. Then, the specification started to be more officially maintained until today. Following the notation used in \cite{Ara2020}, we will refer to the last informal specification (2018) as v0.5.
The current version of the specification is v1.0, and it is the frozen specification for public review. Together with V, it also describes various other extensions, like the ones targeting embedded processors: Zve. Throughout this work, we will only refer to the main extension for application processors. 
Even if the core concept of the \gls{RVV} extension remained the same through time, there have been notable modifications: 1) the organization of the vector register file, 2) the encoding of the instructions, and 3) the organization of the mask registers. We will discuss these major changes in the following subsections.

\subsection{Vector Register File}

\subsubsection{VRF state} The \gls{VRF} is the most critical part in the design of a vector processor. It contains the vector elements, and its layout highly impacts the design choices. When the supported vector length is wide enough, it is usually implemented with \glspl{SRAM}, virtually creating another level in the memory hierarchy.

v0.5: 
The state of the register file was kept both globally and locally. The user dynamically specified how many registers were enabled, and the hardware calculated the maximum vector length by dividing the byte space of the register file among all the enabled registers. Then, each register could be individually programmed to store a different data type.

v1.0:
The state of the register file is only global. The vector register file is composed of 32 VLEN-bit vector registers, where VLEN is an implementation-dependent parameter and indicates the number of bits of a single vector register. It is possible to tune the parameter LMUL to change the granularity of the register file, \eg setting LMUL to $2$ means that the register file will be composed of sixteen $2\times\text{VLEN}$-bit vector registers. Moreover, the register file is agnostic on the data type of the stored elements.

\subsubsection{Striping distance} The original proposal did not constrain the vector register file byte layout in a clear way. Later, the \gls{SLEN} parameter was added to further specify how implementations could organize their internal vector register file byte layout. This parameter became lanes-friendly, especially in version 0.9 of the specifications.

v0.9: 
$\text{SLEN} \le \text{VLEN}$: each vector register is divided into a total of $\text{VLEN}/\text{SLEN}$ sections with SLEN bits. Consecutive vector elements are mapped into consecutive sections, wrapping back around to
the first section until the vector register is full \cite{rvv-0.9}.

v1.0:
$\text{SLEN} = \text{VLEN}$: the \gls{VRF} is seen as a contiguous entity, and consecutive element bytes are stored in consecutive \gls{VRF} bytes.

\subsection{Instruction Encoding}

\hspace{3.5mm}v0.5:
Since the data type of the vector elements was specified in a control register for each vector register, the instructions could use a polymorphic encoding, \eg the instruction \texttt{vadd} would be used to add two vector registers, regardless of their data type.

v1.0:
The encoding is monomorphic, and there are different instructions for different data types, \ie integer, fixed-point, floating-point. The \gls{ISA} has, therefore, more instructions, being one the longest extensions in the whole RISC-V environment.

\subsection{Mask Register Layout}
Mask bits are used to support predication, the way in which vector processors run conditional code. There is one mask bit per element in a vector, and the core executes the instruction on element $i$ only if the $i$-th mask bit has a specific value. 

v0.5:
Only one vector register (\texttt{v1}) could host the mask vector. Each element of this vector could host one mask bit in its \gls{LSB}.

v1.0:
Every register of the \gls{VRF} can be a mask register, and the mask bits are sequentially packed one bit after each other, starting from the \gls{LSB} of the vector register file.


\section{RISC-V V and Lanes}

In this section, we discuss the impact that the RVV extension has on the microarchitecture. We will consider Ara as an example of a design tuned to \gls{RVV} 0.5, even if the discussion is not limited to it. In the following, we refer to it as \gls{VU0.5}, and to our new architecture as \gls{VU1.0}. 

\gls{VU1.0} is a flexible architecture with parametric VLEN developed to obtain high performance and efficiency on a vast range of vector lengths. For example, with $\text{VLEN} = 4096$, the unit can process vectors up to \SI{4}{\kibi\byte}, when $\text{LMUL} = 8$, with a \SI{16}{\kibi\byte} \gls{VRF}. 
Pushing for high vector lengths has many advantages: operating on vectors that do not fit the \gls{VRF} requires strip-mining with its related code overhead, which translates into higher bandwidth on the instruction memory side and more dynamic energy spent on decoding and starting the processing of the additional vector instructions. 

\subsection{VRF and Lanes}
In \gls{VU0.5}, the \gls{VRF} was implemented by splitting it into chunks, one per lane. \gls{VU1.0} maintains the same lane-based \gls{VRF} organization. In this section, we discuss an alternative form of re-implementing the \gls{VRF} with a monolithic architecture, which would increase the routing complexity to/from each bank and the interconnect area, adding an additional dependency on the number of lanes.
In general, the area of the \gls{VRF} crossbar on the ports of the banks ($A_{\text{xbar}}$) is proportional to both the number of masters and slaves, as it requires \texttt{Masters} de-multiplexers and \texttt{Slaves} arbiters.

More in detail, in the lane-based organization, each lane contains a \gls{VRF} section with of 8 1RW \gls{SRAM} banks. All the masters of a lane ($M_{\text{lane}}$) connect to a bank ($B$), so the total interconnect area is the interconnect area of one lane multiplied by the number of lanes ($\ell$),
\begin{equation}
    A_{\text{xbar}}^{\text{split}} \propto M_{\text{lane}} \times \text{B}_{\text{lane}} \times \ell = M_{\text{lane}} \times 8 \times \ell,
\end{equation}
while a monolithic \gls{VRF} would connect each bank to every master of each lane,
\begin{equation}
    A_{\text{xbar}}^{\text{mono}} \propto (M_{\text{lane}} \times \ell) \times B_{\text{tot}} = M_{\text{lane}} \times 8 \times \ell^2.
\end{equation}

The quadratic dependency on the number of lanes would easily limit the potential scaling of a vector processor with a monolithic \gls{VRF}. Moreover, a split \gls{VRF} allows for additional freedom on the macro placement during the floorplanning of the vector unit and improved routability since the interconnect is local to the lane.

\subsection{Byte Layout}

During vector memory operations, the vector processor maps bytes from the memory to bytes in its vector register file. 
Following RVV 1.0, the memory and \gls{VRF} layout must be the same, \ie the $i$-th byte of the vector in memory is stored in byte $i$-th of the \gls{VRF}. 
This condition cannot hold in the case of a split vector register file, as subsequent elements should be mapped to consecutive lanes to better exploit \gls{DLP} \new{and not to complicate mixed-width operations}. 
Since the element width can be changed and the mapping between elements and lanes remains constant, the one between bytes and lanes does not. Depending on the element width, the same byte is mapped to different lanes. 

As a consequence, the processor must keep track of the element width of each vector register to be able to restore its content, and each unit that accesses a whole vector register must be able to remap its elements.

\subsection{Shuffle/Deshuffle} 
The remapping is realized with shuffling (bytes to \gls{VRF}) and deshuffling (bytes from \gls{VRF}) circuits, which translate into a level of byte multiplexers, one per output byte. 
If $N$ lanes operate on a \SI{8}{\byte} datapath, and the units that access the whole \gls{VRF} gather data from each lane in parallel to sustain the throughput requirements, $N \times 8$ bytes are shuffled/deshuffled every cycle, using $N \times 8$ multiplexers. Since an \gls{RVV} unit should support four different element widths (\SIlist{8;16;32;64}{\bit}), each multiplexer has four input bytes.

\subsection{\gls{RVV} 1.0 Implementation Challenges}
Some of the changes introduced by \gls{RVV} 1.0 simplify the interface to software programmers but complicate the design of a vector machine partitioned into lanes. 

\subsubsection{Mask Unit}
When a vector operation is masked, the lane should not update the masked element bytes in the destination vector register. To do so, the information about the masked indices should be available within each lane. Due to the new mask register layout, every vector register can be used and read as a mask register, and because of the new mask vector layout, a lane can need mask bits stored in a different lane. Since data is shuffled within each register of the \gls{VRF}, we need a unit (Mask Unit) capable of fetching and deshuffling the data with the knowledge of the previous encoding used for that register and then expanding and forwarding the masks to the correct lanes.
The introduction of another unit that access all the lanes leads to a greater routing complexity, especially when scaling up the number of lanes, as already noticed in~\cite{Ara2020}.  

\subsubsection{Reshuffle}
The choice of constraining the \gls{VRF} byte layout imposing $\text{SLEN} = \text{VLEN}$ in architectures with lanes leads to peculiar issues when executing specific instruction patterns.
Following the specifications, the architecture should also support the tail-undisturbed policy, \ie the elements past the last active one should not be modified.
When an instruction writes a vector register \texttt{vd} that was encoded with \eew{vd}{old}, with \eew{vd}{new} $\ne$ \eew{vd}{old}, and the old content of the register is not fully overwritten, the previous data get corrupted since the byte mapping of \texttt{vd} is no more unique.

Not to corrupt tail elements, \gls{VU1.0} must deshuffle the destination register using \eew{vd}{old} and reshuffle it back using \eew{vd}{new}. This operation is done by the slide unit since it can access all the lanes and already has the necessary logic to perform this operation (which is a \texttt{vslide} with null stride, and different \eew{}{} for the source and destination register). We called this operation \textit{reshuffle}. 

The issue is exacerbated by the fact that the program can change the vector length and element width at runtime, so it is not possible to know how many bytes need to be reshuffled unless both the vector length and the element widths are dynamically tracked for each vector register. Without this knowledge, the architecture must always reshuffle the whole register. In our architecture, the vector unit injects a reshuffle operation as soon as the front-end decodes an instruction that writes a vector register changing its encoded \eew{}{}. The reshuffle is executed before the offending instruction, and it is not injected if the instruction writes the whole vector register.
In general, reshuffling hurts the \gls{IPC} if the latency of the shuffle cannot be hidden and if this operation causes structural hazards on following slide instructions.
A special case of instructions that suffer from this problem are the narrowing instructions, especially those that have the same source and destination registers. Without a renaming stage in the architecture, it is not possible to decouple source and destination registers, and a simple reshuffle operation is not enough to preserve the tail elements: when reshuffling to a lower EEW, bytes that belong to the same element are placed in different lanes, and cannot, therefore, be fetched anymore within one lane by the narrowing operation.

Using tail-agnostic policies presents some issues as well. The tail bytes must be either left unchanged or overwritten with ones. The bytes that are left unchanged get corrupted since the information about their mapping also gets lost, and writing all the tail elements to 1 has significant overhead, as these additional writes deteriorate the \gls{IPC} and possibly generate new bank conflicts on the \gls{VRF}.

Reshuffling is a costly operation, as the offending instruction always has a \gls{RAW} dependency on the reshuffle: the cost is even higher if the throughput of the slide unit is lower than the one of the computational unit, as chaining cannot proceed at full speed.
The compiler can alleviate the problem by clustering the vector register file avoiding change in EEW in the same register as much as possible. On the hardware side, a renaming stage could also help in prioritizing the remapping of the destination registers to physical registers with the same EEW.  


\begin{figure*}
    \centering
    \includegraphics[width=\linewidth]{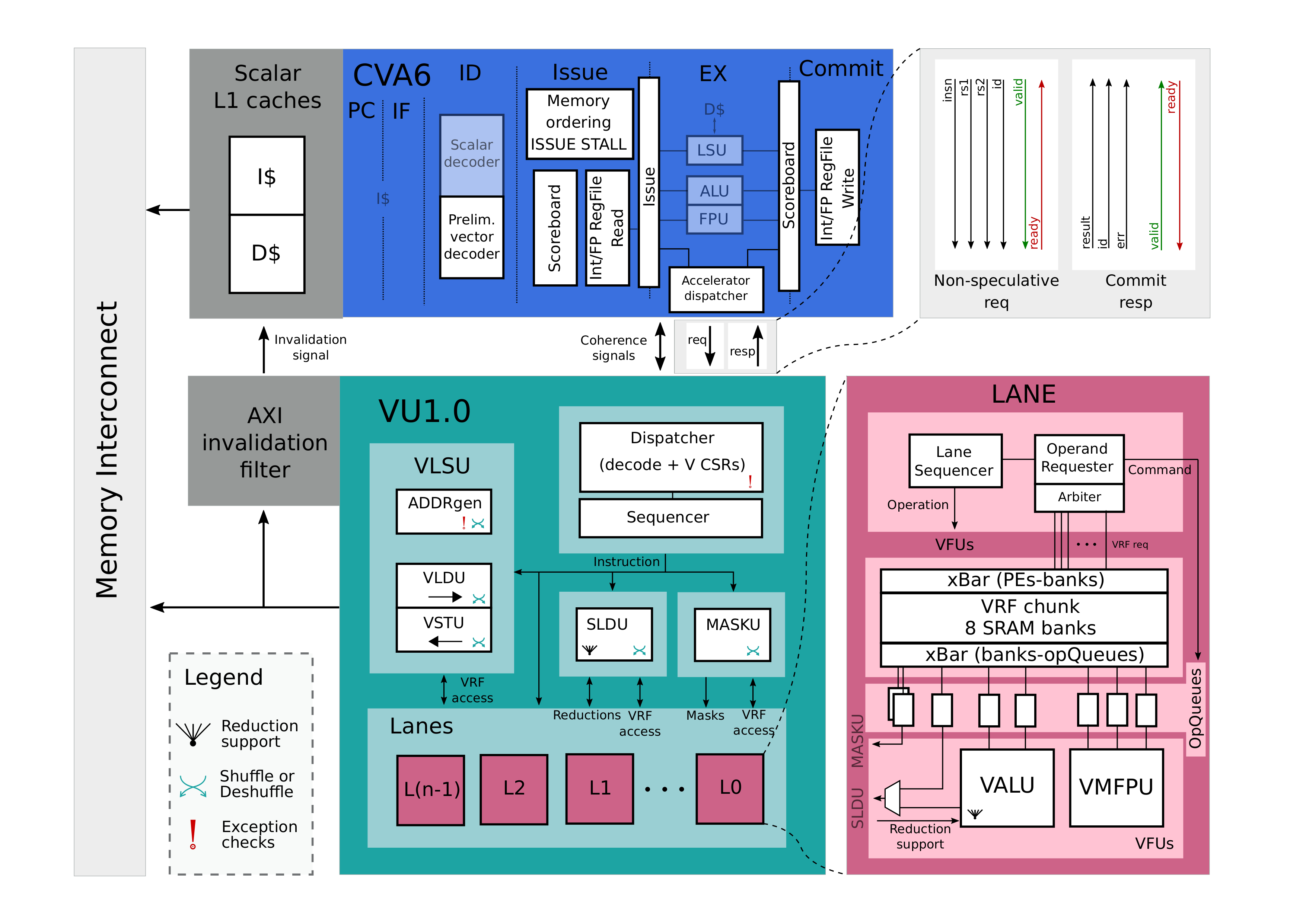}
    \caption{Top-Level block diagram of (the new) system with the vector co-processor marked in green, a more detailed diagram of the lane in magenta, and the host scalar core CVA6 in blue.}
    \label{fig:main-dg}
\end{figure*}

\section{Architecture}
Our unit supports the vast majority of \gls{RVV} 1.0 with the following exceptions: no support for fixed-point arithmetic, floating-point reductions and very specific instructions (round towards odd, reciprocal, square root reciprocal), segment memory operations (non-mandatory since \gls{RVV} 1.0), gather-compress, scalar moves (we emulate them through memory transfers), and some specific mask instructions (e.g., \texttt{vfirst}, \texttt{viota}).
\camrdy{In Figure~\ref{fig:main-dg}, we show the main block diagram of the system.}
The integration with the CVA6 processor is based on the RVV0.5 unit \cite{Ara2020}, with the following major enhancements.

\paragraph{Decoding}
With the new RISC-V V specifications, the encoding of the vector instructions now fully specifies the data type of the vector elements on which the instruction operates. This allows moving most of the decoding logic and vector-specific \glspl{CSR} from CVA6 to the vector unit, making CVA6 more agnostic on the V extension.
In the updated architecture, CVA6 keeps only the pre-decoding logic strictly needed to know 1) if the vector instruction is a vector instruction, to dispatch it to the vector unit when it reaches the head of the scoreboard, 2) if the vector instruction is a memory operation (needed for cache coherency), and 3) if the instruction needs a scalar value from the integer or floating-point scalar register files.

\paragraph{CVA6-Vector Unit Interface}
The interface between the host processor CVA6 and the vector unit is generalized: the unit is implemented as a modular accelerator with its own CSR file. While decoding, CVA6 identifies vector instructions, pushes them to a dispatcher queue, and dispatches them to the accelerator once they are no longer speculative.

\paragraph{Memory Coherency}
CVA6 and the vector unit feature separate memory ports, and CVA6 has a private L1 data cache.
At the same time, the \mbox{RISC-V} ISA mandates a strictly coherent memory view between the scalar and vector processors.
\gls{VU0.5}~\cite{Ara2020} violates this requirement and needs explicit memory fences that write back and invalidate the CVA6 data cache between accesses on shared memory regions, adding a significant performance overhead and reducing code portability. In our \gls{VU1.0}, we extend the system by a lightweight hardware mechanism to ensure coherency. We adapt the CVA6 L1 data cache to a write-through policy so that the main memory, which is accessed by the vector unit as well, is always up-to-date. When the vector unit performs a vector store, it invalidates the corresponding cache lines in the CVA6 data cache. Moreover, we issue 1) scalar loads only if no vector stores are in-flight, 2) scalar stores only if no vector loads or stores are in-flight, and 3) vector loads or stores only if no scalar stores are pending.

\paragraph{Mask Unit}

After the update, mask bits are not always in the correct lane. Thus, we design a Mask Unit, which can access all the lanes at once to fetch, unpack and dispatch the correct mask bits to the corresponding lanes.

\paragraph{Reductions}
Since our design has lanes, we implement integer reductions using a 3-steps algorithm: intra-lane, inter-lanes, and SIMD reduction steps. The intra-lane reduction fully exploits the data locality within each lane, maximizing the ALU utilization and efficiency, reducing all the elements already present in the lane. The inter-lanes reduction moves and reduces data among the lanes in $\log_2(\ell)+1$ steps, $\ell$ being the number of lanes, using the slide unit; since there is a dependency feedback between the slide and ALU units, the latency overhead of the communication is paid at every step. Finally, the \gls{SIMD} reduction reduces the \gls{SIMD} word, if needed; therefore, its latency logarithmically depends on the element width.

\section{Results}
\label{sec:results}
One of the main motivations under the design of a vector processor is to efficiently maximize the throughput exploiting the intrinsic application \gls{DLP}. In the following sections, we will use the word \textit{throughput} to indicate the number of useful computational results produced per clock cycle, \eg when adding two vectors of length $N$, the throughput is $N$/\#cycles, where \#cycles is the number of cycles required to produce the $N$ results. 
With our experiment, we explore how the changes introduced by the updated V specification and the novel features of the system affect the throughput.
Moreover, we place and route our design and extract the related \gls{PPA} metrics.

\subsection{Performance}
\label{ssec:performance}
We manually optimize the benchmarks used in \cite{Ara2020} (fmatmul, fconv2d with $7 \times 7 \times 3$ kernel), adapting them to the new specifications and architecture, and compile them using LLVM 13.0.0. To make a comparison between the two systems, we measure the cycle count of the same benchmarks of~\cite{Ara2020} with a cycle-accurate simulation of our vector unit using Verilator v4.214. We tune the benchmarks in assembly, as the LLVM compiled code showed inferior performance w.r.t.\ the optimized one.

To provide further insights on the system, we run the same experiment measuring how much the scalar core limits the final performance because of its non-ideal issue rate of vector instructions to the vector unit, showing how the scalar memory accesses impact the final throughput.

\begin{figure}
  \centering

    \resizebox{1\linewidth}{!}{
  \begin{tikzpicture}
    \begin{axis}[
      xlabel = {Matrix size $n$ [elements]},
      log basis x = {2},
      xmode = log,
      xmin = 8,
      xmax = 128,
      ylabel={Performance [\si{\double\flop\per\cycle}]},
      log basis y = {2},
      ymode = log,
      ymin = 1,
      ymax = 40,
      log ticks with fixed point,
      grid = major,
      legend style = {at={(1,0)}, anchor=south east, legend columns=-1, font=\small}]
      
      \addlegendimage{line legend, thick, color1}
      \addlegendimage{line legend, thick, color2}
      \addlegendimage{line legend, thick, color3}
      \addlegendimage{line legend, thick, color4}

      \draw[fill=color1, opacity=0, fill opacity=.15]
      (axis cs:4,1) -- (axis cs:128,1) -- (axis cs: 128,4) -- (axis cs: 4,4) -- (axis cs: 4,1);

      \draw[fill=color2, opacity=0, fill opacity=.15]
      (axis cs:4,4) -- (axis cs:128,4) -- (axis cs: 128,8) -- (axis cs: 4,8) -- (axis cs: 4,4);

      \draw[fill=color3, opacity=0, fill opacity=.15]
      (axis cs:4,8) -- (axis cs:128,8) -- (axis cs: 128,16) -- (axis cs: 4,16) -- (axis cs: 4,8);

      \draw[fill=color4, opacity=0, fill opacity=.15]
      (axis cs:4,16) -- (axis cs:128,16) -- (axis cs: 128,32) -- (axis cs: 4,32) -- (axis cs: 4,16);

      \addplot[dashed, color1, thick, domain=0.25:128, samples=201, name path=roof2]{roof(x,8,4)};
      \addlegendentry{$\ell = 2$}

      \addplot[dashed, color2, thick, domain=0.25:128, samples=201, name path=roof4]{roof(x,16,8)};
      \addlegendentry{$\ell = 4$}

      \addplot[dashed, color3, thick, domain=0.25:128, samples=201, name path=roof8]{roof(x,32,16)};
      \addlegendentry{$\ell = 8$}

      \addplot[dashed, color4, thick, domain=0.25:128, samples=201, name path=roof16]{roof(x,64,32)};
      \addlegendentry{$\ell = 16$}

      \addplot [mark=square, color1] table {tables/matmul_2l};
      \addplot [mark=square, color2] table {tables/matmul_4l};
      \addplot [mark=square, color3] table {tables/matmul_8l};
      \addplot [mark=square, color4] table {tables/matmul_16l};
      
      \addplot [dotted, thick, color5, domain=8:64] {x/2};

      \addplot [thick, mark=square*, color1] table [x index=0, y index=2] {tables/matmul_2l};
      \addplot [thick, mark=square*, color2] table [x index=0, y index=2] {tables/matmul_4l};
      \addplot [thick, mark=square*, color3] table [x index=0, y index=2] {tables/matmul_8l};
      \addplot [thick, mark=square*, color4] table [x index=0, y index=2] {tables/matmul_16l};
    \end{axis}
  \end{tikzpicture}
  }
  \caption{Runtime of matrix multiplication kernel of size $n \times n$ on our CVA6+Vector Unit system ($\blacksquare$), compared with the ideal dispatcher ($\square$), for several number of lanes $\ell$.}
  \label{fig:ideal-runtime-matmul}
\end{figure}
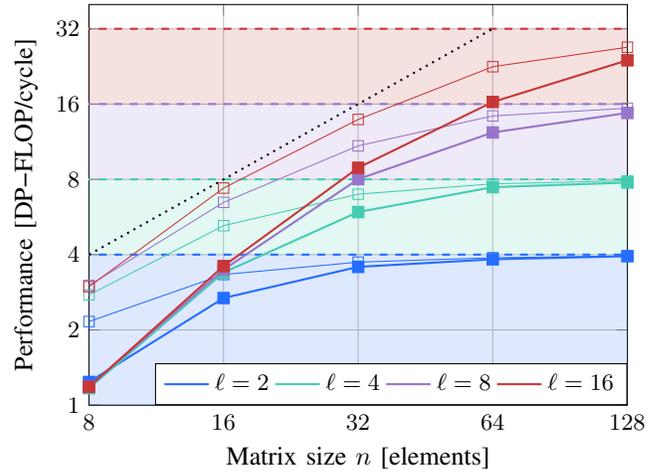

\Cref{fig:ideal-runtime-matmul} shows Ara's roofline model for a variable number of lanes, and the performance results of the matrix multiplication benchmark between square matrices $n \times n$ for several matrix sizes $n$. The arithmetic intensity for this benchmark is proportional to $n$. The horizontal dashed lines mark the computational limit of the architecture for the corresponding number of lanes. 
Originally, Ara achieved almost peak performance on the com\-pu\-te-bound \texttt{fmatmul} and \texttt{fconv2d} kernels, showing high levels of \glspl{FPU} utilization; this was also shown by Hwacha, with utilization above 95\% \cite{hwachav4}. Despite having a \gls{VRF} 1/4 of the size of \cite{Ara2020}, our new architecture achieves comparable or better performance for long vectors both for \texttt{fmatmul} and \texttt{fconv2d}, with a peak of utilization of >98.5\% with 2 lanes on $128 \times 128$ \texttt{fmatmul}. Since the utilization is almost 100\%, an increase in \gls{VRF} size would barely increase performance for larger problems.

In \Cref{fig:ideal-runtime-matmul}, we show both the performance of the real system and that extracted with a perfect dispatcher. The perfect dispatcher is simulated by replacing CVA6 with a pre-filled queue with the corresponding vector instructions. The system with the ideal dispatcher shows the real performance limitations of the vector architecture and marks a hard limit on the performance achievable only by optimizing the scalar part of the system itself.
The dotted black diagonal line is the limit given by the issue rate of computational vector instructions. 
In \cite{Ara2020}, the authors identify a hard limit to the performance of the matrix multiplication on the system, especially for short vectors. With \gls{RVV}~v0.5 and their algorithm, the issue rate of computational instructions for the main kernel is one instruction every five cycles. This was due to the presence of the \texttt{vins} instruction, used to move a scalar value from CVA6 to Ara. With the new specification, this is no longer needed since scalar operands can be passed with the vector multiply-accumulate instruction. This improves the computational instruction issue rate limit from 1/5 to 1/4, shifting the diagonal line to the left.

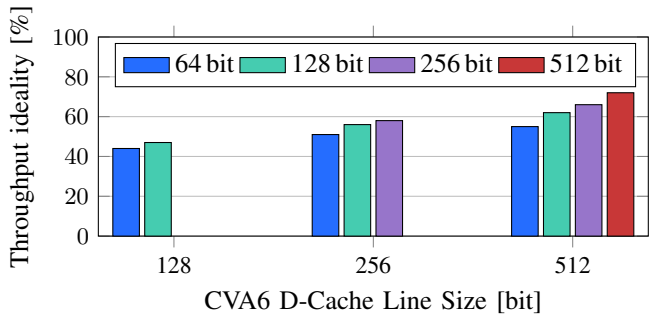
\begin{figure}
    \centering
    \resizebox{1\linewidth}{!}{
    \begin{tikzpicture}
        \begin{axis}[
        ybar,
        ymin = 0,
        ymax = 100,
        xlabel = {CVA6 D-Cache Line Size [\si{\bit}]},
        ylabel = {Throughput ideality [\si{\percent}]},
        enlarge x limits={abs=1cm},
        xtick = data,
        symbolic x coords = {128, 256, 512},
        legend pos = {north west},
        legend columns=-1,
        height = 120,
        ymajorgrids]
            \addplot[ybar, fill=color1, area legend] coordinates {(128,44) (256, 51) (512, 55)};
            \addplot[ybar, fill=color2, area legend] coordinates {(128,47) (256, 56) (512, 62)};
            \addplot[ybar, fill=color3, area legend] coordinates {         (256, 58) (512, 66)};
            \addplot[ybar, fill=color4, area legend] coordinates {                   (512, 72)};
        
            \legend{\SI{64}{\bit}, \SI{128}{\bit}, \SI{256}{\bit}, \SI{512}{\bit}}
        \end{axis}
    \end{tikzpicture}
    }
    \caption{System throughput ideality relative to system with ideal dispatcher, as a function of CVA6's D-cache line size and \gls{AXI} data width.}
    \label{fig:scalar_ideality}
\end{figure}

To study the impact that the scalar part of the system has on performance, we modify the \gls{AXI} data width of the memory port of CVA6, as well as the D-cache line width, impacting the miss-penalty and miss-rate of the scalar data memory requests and, therefore, influencing the issue rate of the vector instructions of a kernel that, at every iteration, needs new scalar elements that are forwarded to the vector unit.
In Figure~\ref{fig:scalar_ideality}, we summarize the throughput ideality of a 16 lane system executing \texttt{fmatmul} on a $16 \times 16$ matrix when varying the memory parameters of the scalar core. Increasing the cache line size decreases the miss rate, but if this comes without widening the \gls{AXI} data width, the miss penalty is negatively influenced. The system throughput when both the cache line and \gls{AXI} data width are 512-bit is $1.54\times$ larger than when they are both set to $128$ bits, showing the importance of the sizing of the scalar memory part of the system when improving the performance on medium/short vectors.

\paragraph{Short vectors} Even if the issue-rate limit is now improved for this kernel thanks to the new specification, short vectors' performance is lower than ideal. Our \gls{VRF} does not implement the barber-pole VRF layout; so, the number of effective banks in each lane is reduced, and the system experiences more bank conflicts. For example, with 16 elements on 16 lanes, every element is stored in bank 0 in each lane, and every read/write operation will target the same bank. In general, if there is not at least one element per bank (128 elements for 16 lanes, with $\text{VLEN} = 4096$), there will be more bank conflicts and related stalls. Barber-pole can mitigate this issue, as also with 16 elements in 16 lanes, elements of different registers occupy different banks.
This is not a critical issue, as our vector architecture primarily targets long vectors; moreover, low performance on short vectors was already observed in Hwacha and Ara, even when implementing Barber-Pole layout in their \gls{VRF} \cite{Ara2020}. Designers can choose a lower vector length or opt for more efficient non-vector \gls{SIMD} architectures.

\paragraph{Reductions}
In \tabref{tab:tabred} we report the performance and efficiency results from running the dot product kernel, varying the number of lanes, the vector byte length, and the element width. The measured cycle count only refers to the actual computational dot product, \ie the vector-vector element-wise multiplication and the subsequent reduction, without the memory operations. Since our unit's multiplier and adder belong to different functional units, the product and reduction can be successfully chained so that the final cycle count scales only with the number of elements in the vector and not with the number of instructions (in this case, two). The efficiency is calculated on the ideal performance counted as $\text{VL}_\text{B}/8\ell + 1 + \log_2(\ell)$, where $\text{VL}_\text{B}$ is the vector length in bytes, and the added $1$ keeps into account the chaining of the multiplication.
1) Longer vectors linearly increase the execution time of the intra-lane reduction. The longer the vector, the more this step hides the overhead of the other two, contributing to higher efficiencies.
2) The number of lanes linearly speeds up the intra-lane reduction phase and logarithmically negatively impacts the time spent during the inter-lanes reduction. When this phase is not hidden enough (\eg short vectors), the overhead can be important w.r.t. the theoretical maximum achievable. The more the lanes, the \textit{shorter} is the vector, in relative terms. To reach high levels of efficiency, a design with more lanes requires longer vectors. 
3) Lower element widths positively influence the throughput, adding only a logarithmic overhead during the SIMD-reduction phase.
The advantage over the scalar core is critical and can lead up to $380\times$ of performance improvement, especially for reduced element widths and long vectors, where the scalar cycle count skyrockets (\textgreater24k cycles peak): our vector unit exploits SIMD-like computation to keep a constant execution time while the element width is lowered, with a negligible overhead during the SIMD reduction phase.

For short vectors, the startup time of the vector operations is not amortized, and the overall efficiency drops. For example, our vector unit needs about ten cycles to practically produce results from the reduction after the vector multiplication is issued to the back-end.

\begin{table}
\caption{
Cycle count for reduction operation, with 2/16 lanes, vector byte-lengths, and vector element sizes. Slashes divide results for different element sizes: 8-bit elements (left), and 64-bit elements (right).}
\label{tab:tabred}
\scriptsize
\begin{tabular}{r@{\hspace{2mm}}c@{\hspace{2mm}}c@{\hspace{2mm}}c@{\hspace{2mm}}c@{\hspace{2mm}}c@{\hspace{2mm}}c}
\toprule
 & \multicolumn{3}{c}{\textbf{Cycle Count (\#)}}  & \multicolumn{3}{c}{\textbf{Efficiency (\% on ideal)}}  \\ 
\textbf{}         & \textbf{64 B} & \textbf{512 B} & \textbf{4096 B} & \textbf{64 B} & \textbf{512 B} & \textbf{4096 B} \\ \midrule
\textbf{2 Lanes}  & 25 / 23       & 55 / 51        & 279 / 275       & 24\% / 26\%   & 62\% / 67\%    & 92\% / 94\% \\ 
\textbf{16 Lanes} & 33 / 32       & 36 / 32        & 64 / 60         & 17\% / 17\%    & 25\% / 28\%    & 58\% / 62\% \\ 
\bottomrule
\end{tabular}
\end{table}

\subsection{Physical Implementation}
To assess the impact of our architectural modifications on the \gls{PPA} metrics of the system, we synthesize and place-and-route our 4-lane enhanced design (CVA6, its caches, and the new vector unit) with $\text{VLEN} = 4096$ (\SI{16}{\kibi\byte} of \gls{VRF}), targeting \textsc{GlobalFoundries} 22FDX \gls{FDSOI} technology. \new{The scalar I-cache and the D-cache have a line-width of 128 and 256 bits, respectively.}
We use Synopsys DC Compiler 2021.06 for synthesis with topographical information and Synopsys IC Compiler II 2021.06 for the physical implementation.
We extract post-layout area and frequency, and, lastly, the related power results by means of activity-based power simulations \new{with SDF back-annotation} on typical conditions, elaborated using Mentor QuestaSim 2021.2 and Synopsys PrimeTime 2020.09\new{, while executing a $128 \times 128$ \texttt{fmatmul}}.

\begin{table}
\caption{Physical implementation comparison between VU0.5 and VU1.0}
\label{tab:system-qor}
\scriptsize
\begin{tabular*}{\linewidth}{@{}l@{\hspace{1mm}}r@{\hspace{1mm}}r@{\hspace{1mm}}r}
\toprule
\textbf{} & \textbf{VU0.5 System} & \textbf{VU1.0 System} & \textbf{Update Merit} \\
\midrule
\textbf{VRF Size [\SI{}{\kibi\byte}]} & 64 & 16 & -75\% \\ 
\textbf{Die Area [\SI{}{\square\milli\meter}]} & 0.98 & 0.81 & -15\% \\ 
\textbf{Cell Area [\SI{}{\square\milli\meter}]} & 0.43 & 0.49 & +14\% \\ 
\textbf{Memory macro Area [\SI{}{\square\milli\meter}]} & - & 0.15 & - \\
\textbf{Worst case frequency [\SI{}{\mega\hertz}]} & 925 & 920 & -0.5\% \\ 
\textbf{TT frequency [\SI{}{\giga\hertz}]} & 1.25 & 1.34 & +7.2\% \\ 
\textbf{Performance [DP-GFLOPS]} & 9.8 & 10.4 & +6.1\% \\ 
\textbf{Power @TT frequency [\SI{}{\milli\watt}]} & 259 & 280 & - \\ 
\textbf{Efficiency [DP-GFLOPS/W]} & 37.8 & 37.1 & -1.9\% \\ 
\bottomrule
\end{tabular*}
\end{table}

\begin{figure}
    \centering
    \includegraphics[width=\linewidth]{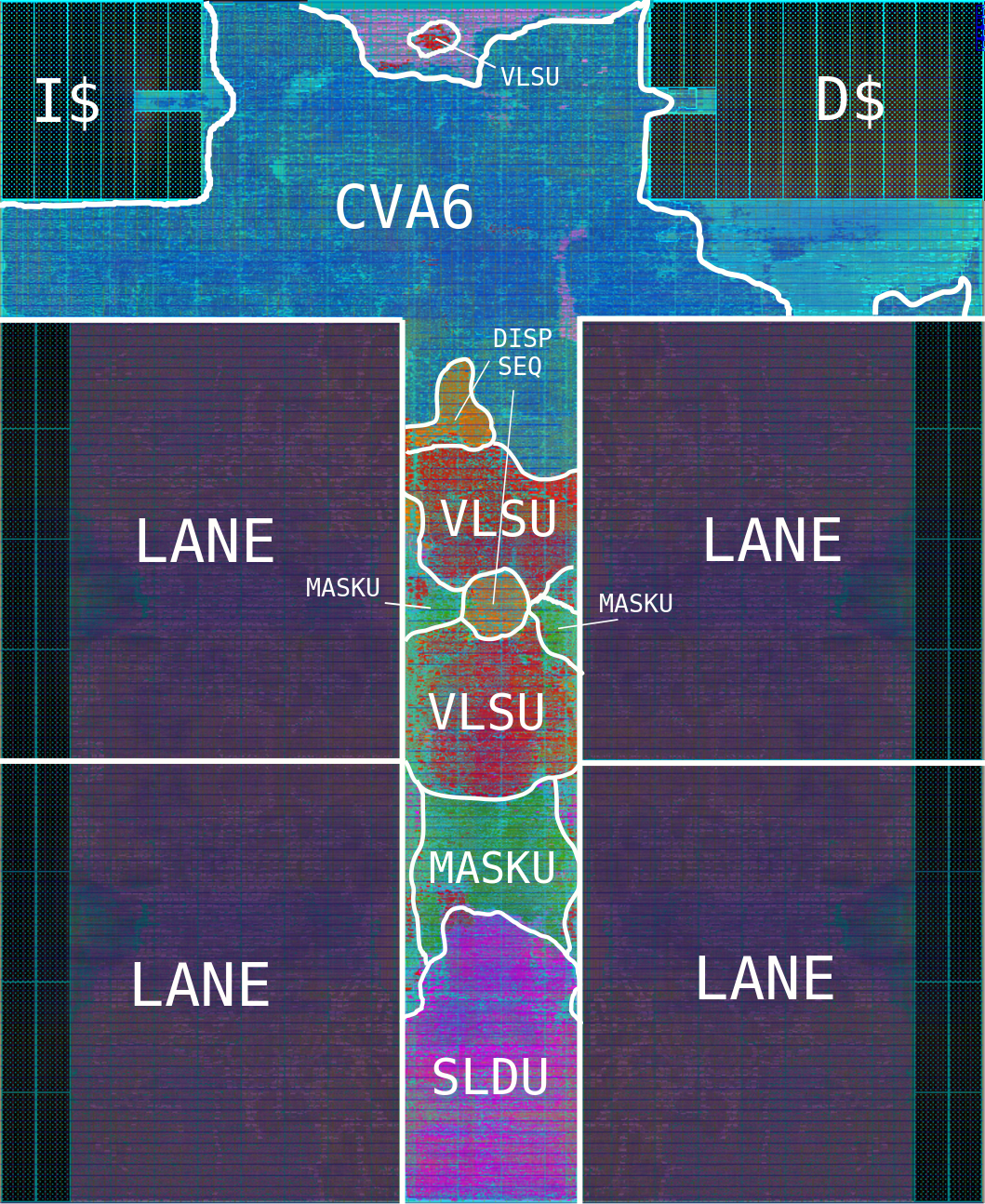}
    \caption{Physical implementation of the full system. The lane is implemented and enclosed in a macro and then placed on the die. The system input and output are at the top of the die (AXI interface).}
    \label{fig:system-pnr}
\end{figure}

\begin{figure}
    \centering
    \includegraphics[width=\linewidth]{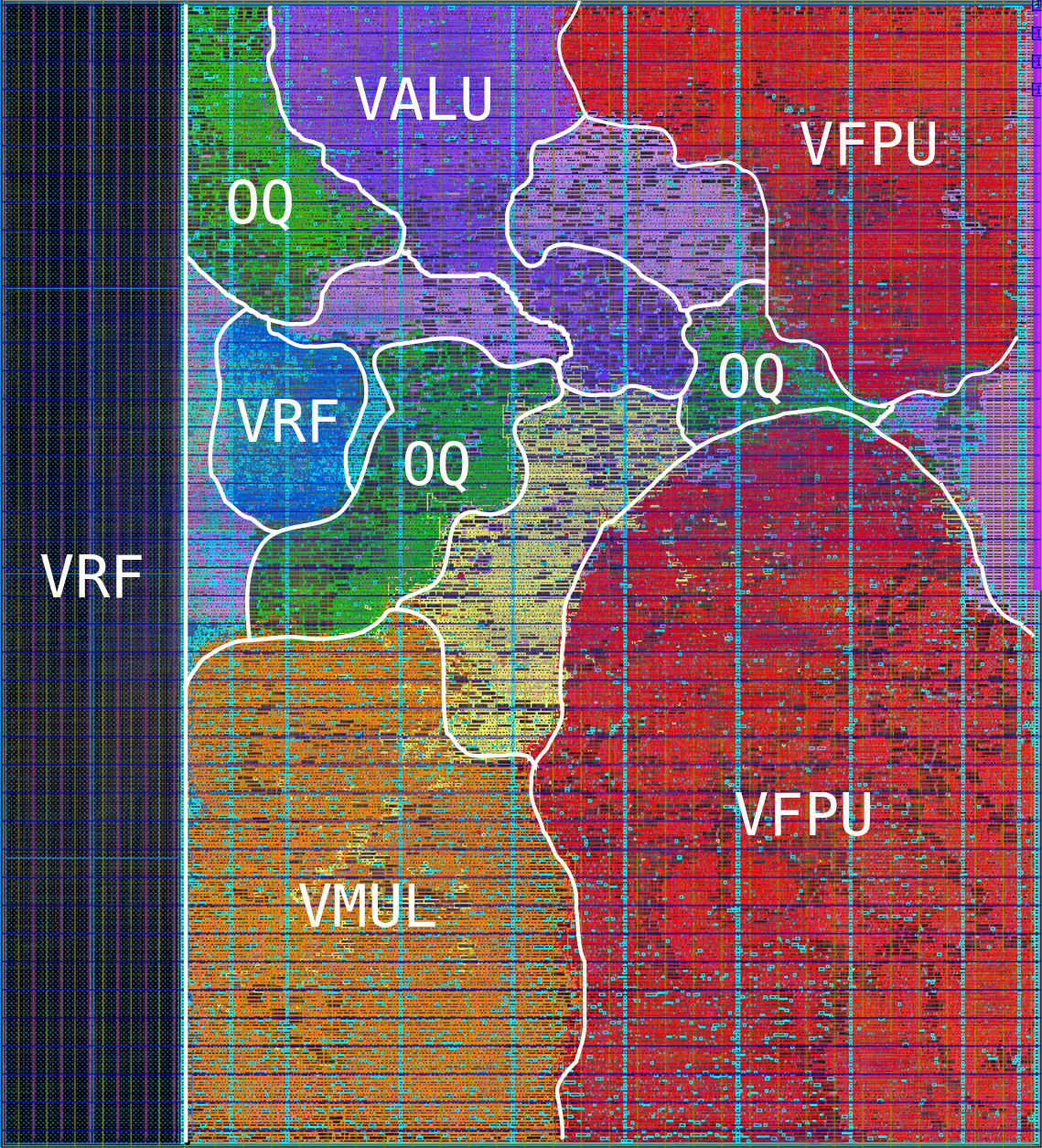}
    \caption{Physical implementation of a Lane. Modules without a label: lane sequencer, operand requesters (close to the \gls{VRF}), \gls{VDIV} and control logic for \gls{VMUL} and \gls{VFPU} (in the middle).}
    \label{fig:lane-pnr}
\end{figure}

The design is placed and routed as a $0.81$ \textit{mm} $\times$ $1.00$ \textit{mm} macro.
To improve the process, we develop an implementation flow that leverages the modularity of the design.
Our vector unit is composed of a parametric number of lanes that contain most of the processing logic of the system. All the lanes are identical, and their synthesis requires automatic retiming of the pipeline stages of the \gls{FPU}. We use a hierarchical synthesis and back-end flow, in which the lanes are designed as custom macros and synthesized once to reduce turn-around time significantly.

\begin{figure}
    \centering
    \includegraphics[width=\linewidth]{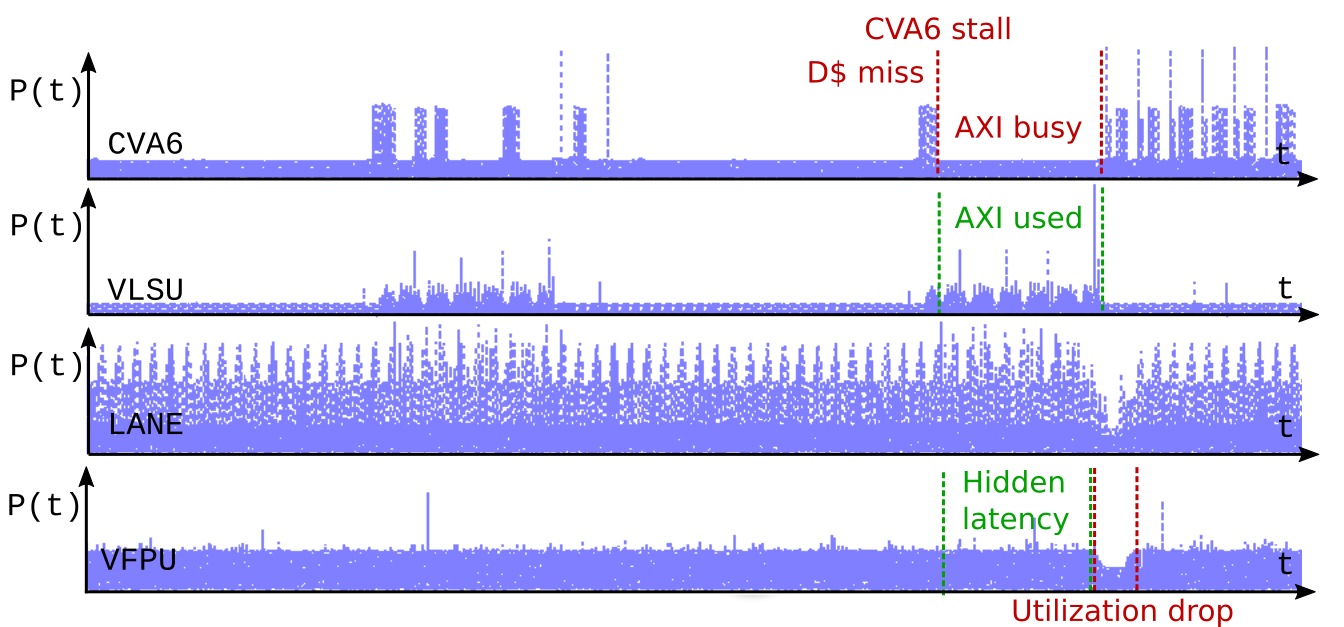}
    \caption{Time-based power simulation - two iterations of \texttt{fmatmul}. VU1.0 performs two vector loads, whose latency is hidden by the computation in its VFPUs (with utilization around 97\%). Towards the end of the time span, CVA6 is temporarily stalled because of an L1D-cache miss that cannot be served by the upper memory system, which is already used by VU1.0's VLSU. To maximize visibility, every sub-plot is scaled to its maximum power consumption.}
    \label{fig:time-based-power}
\end{figure}

\begin{figure}
    \centering
    \includegraphics[width=\linewidth]{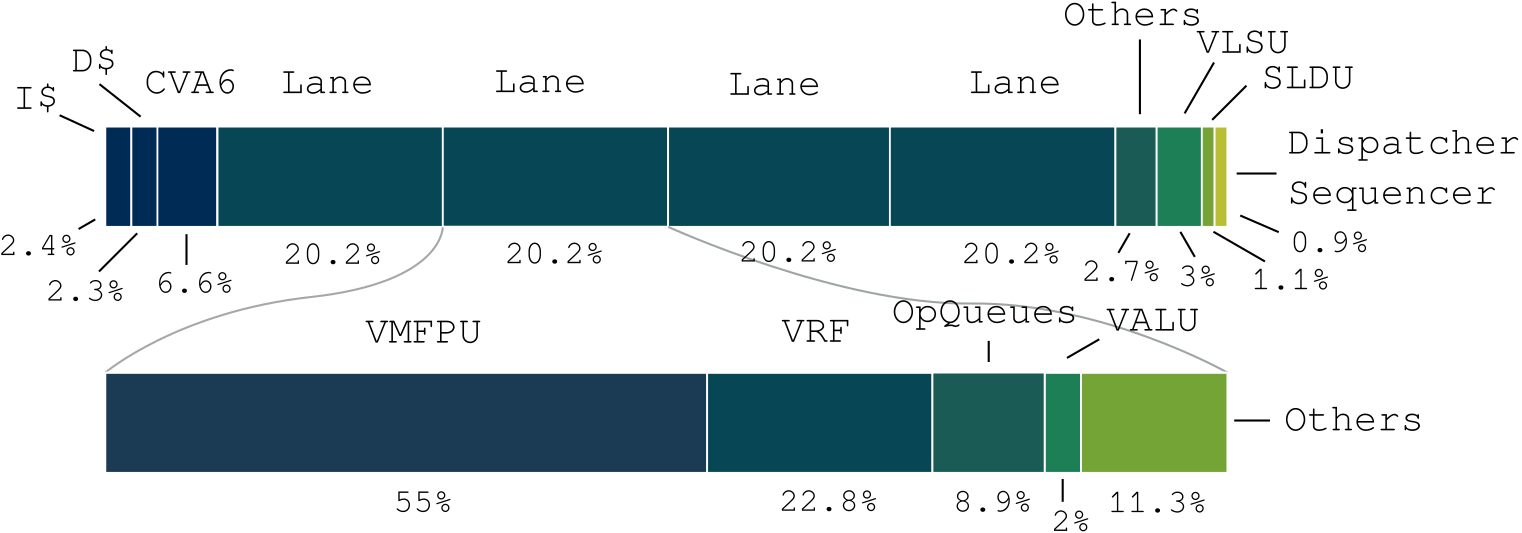}
    \caption{Power breakdown of the system. The overall average power consumption is 280 mW at 1.34 GHz when executing \texttt{fmatmul}.}
    \label{fig:power-breakdown}
\end{figure}

\new{In Figure~\ref{fig:system-pnr} and Figure~\ref{fig:lane-pnr}, we present the physical layout of the whole system and a single lane, respectively. At the top of the die, we have the AXI interface, around which there are the AXI crossbar, CVA6, and part of the \gls{VLSU}. The lanes are on the two sides to ease the routing of all the cross-lane units that access them, like the \gls{MASKU}, the \gls{SLDU}, the \gls{VLSU}, and the main sequencer.}
\new{The lane area is dominated by the \gls{VMFPU}, the module that contains the \gls{FPU} and the \gls{SIMD} multipliers.}

\tabref{tab:system-qor} shows the parameters and the quality figures of our implementation, with respect to \gls{VU0.5}.
Since, as shown in \Cref{ssec:performance}, \gls{VU1.0} achieves a competitive throughput despite using a \gls{VRF} $4\times$ smaller than \gls{VU0.5}'s, we obtain an overall area reduction larger than \SI{15}{\percent} \new{on the die size without trading off performance}. 
\gls{VU1.0} achieves a frequency of \SI{920}{\mega\hertz} in worst-case conditions (SS, \SI{0.72}{\volt}, \SI{125}{\celsius}), virtually the same as in \gls{VU0.5}. \gls{VU1.0} achieves \SI{1.34}{\giga\hertz} in typical conditions (TT, \SI{0.80}{\volt}, \SI{25}{\celsius}), \SI{7.2}{\percent} faster than what is reported in \cite{Ara2020} for \gls{VU0.5}, thanks to an advanced hierarchical implementation strategy. This translates into a \texttt{fmatmul} peak throughput \SI{6.1}{\percent} higher than \gls{VU0.5} (\num{10.4}~DP-GFLOPS). 

\gls{VU1.0} consumes \SI{280}{\milli\watt} while running a $128 \times 128$ \texttt{fmatmul}, which leads to an energy efficiency of \num{37.1}~DP-GFLOPS/W. This is only \SI{1.9}{\percent} lower than \gls{VU0.5}, while supporting a much more complete \gls{ISA}, including reductions and memory coherence support. \neww{In \cite{hwachav5}, the measured efficiency for comparable voltage (\SI{0.8}{\volt}) or frequency (\SI{1.34}{\giga\hertz}) is lower than \num{33}~DP-GFLOPS/W} for a real system running Hwacha 4. \gls{VU1.0}'s efficiency is also much higher than the one reported for Hwacha 4.5~\cite{9567768}, although a direct comparison is not possible, as they only report the power measurements of the full manufactured system.

\new{Figure~\ref{fig:time-based-power} shows the power traces of two iterations of a \texttt{fmatmul} run, highlighting how the vector architecture can tolerate stalls on the scalar core. The vector unit and CVA6 compete for the AXI port to the L2 memory, and, during the second iteration, CVA6 stalls because of an L1 D-cache miss that cannot be served since a vector load is ongoing. During the interval in which CVA6 cannot forward new vector instructions, the vector unit does not starve until it has processed all the elements of the vector in its \gls{VRF}, and the overall utilization remains at its peak for most of the time.}

\new{The average power breakdown of the system running \texttt{fmatmul} is shown in Figure~\ref{fig:power-breakdown}. More than 80\% of the power is consumed by the lanes, where the computation takes place. As expected, the \gls{VMFPU} uses most of the energy of a lane and, together with the \gls{VRF} and the operand queues, accounts for almost 90\% of the total budget of a lane. This highlights how \gls{VU1.0} exploits the regular execution pattern of vector machines, leading to smaller and simpler controller logic. In comparison, CVA6 and its scalar caches consume less than 12\% of the total power budget in spite of the significant area footprint of the scoreboard and instruction dispatch logic~\cite{Zaruba2020}.}

\section{Conclusions}
In this paper, we present the first open-source vector processor compliant with the core of \gls{RVV} 1.0. We compare our design with a \gls{RVV} 0.5 unit and discuss the impact that the specification update has on architectures with a \gls{VRF} split among lanes, the newly added features. We show competitive throughput and \gls{PPA} results using the advanced \textsc{GlobalFoundries} 22FDX technology. The system runs at up to \SI{1.34}{\giga\hertz}, with peak FPU utilization >98\% on crucial matmul kernels. We provide insights on the performance of the mixed scalar-vector system for short vectors and an analysis of the new reduction engine that leads to speedups up to $380\times$ with respect to a scalar core. The source code is released on \url{https://github.com/pulp-platform/ara}.

\section*{Acknowledgment}
This work was supported by the ETH Future Computing Laboratory (EFCL), financed by a gift from Huawei Technologies.

\bibliographystyle{IEEEtran}
\bibliography{ara}
\clearpage

\end{document}